# Cryptanalysis of a Chaotic Key based Image Encryption Scheme

Pratyusa Mukherjee[1], Krishnendu Rarhi (ORCID: 0000-0002-5794-215X)[2], Abhishek Bhattacharya[3]

[1] Indian Institute of Engineering Science and Technology, Shibpur
[2], [3] Institute of Engineering & Management, Kolkata
[1, 2] {pratyusa.mukherjee, rarhik}@gmail.com, [3] abhishek.bhattacharya@iemcal.com

**Abstract.** Security of multimedia data is a major concern due to its widespread transmission over various communication channels. Hence design and study of good image encryption schemes has become a major research topic. During the last few decades, there has been a increasing in chaos-based cryptography. This paper proposes an attack on a recently proposed chaos based image encryption scheme. The cryptosystem under study proceed by first shuffling the original image to disturb the arrangement of pixels by applying a chaotic map several times. Second, a keystream is generated using Chen's chaotic system to mix it with the shuffled pixels to finally obtain the cipher image. A chosen ciphertext attack can be done to recover the system without any knowledge of the key. It simply demands two pairs of plaintext-ciphertext to completely break the cryptosystem.

**Keywords:** Cryptanalysis, Chaos, Keystream, Shuffle, Chosen-ciphertext attack , Image Encryption

## 1   Introduction

With rapid dissemination of digital multimedia data over the Internet, security of digital information is a major concern. Recently, due to advancements in the theory and application of chaos, many researchers are focusing on the chaotic cryptography.

Chaotic maps have attracted the attention of cryptographers as a result of the following fundamental properties [1]:

- Chaotic maps are deterministic, meaning that their behavior is predetermined by mathematical calculations.
- Chaotic maps are unpredictable and nonlinear because they are sensitive to initial conditions. Even a very slight change in the starting point can lead to a significantly different outcome.
- Chaotic maps appear to be random and disorderly but, in fact, they are not; beneath the random behavior is an order and pattern.

These properties can be connected with the "confusion" and "diffusion" property in cryptography. Hence, Chaos are used to enrich the design of new ciphers [3-15]. We are mainly interested on those schemes dedicated to image encryption. Image encryption is a bit different from text encryption due to certain characteristics of an image such as: redundancy of data, a strong correlation among adjacent pixels and a minute change in the attribute of any pixel does not drastically degrade the quality of the image. Some of the encryption schemes have been cryptanalysed and have been found to be insecure. In this paper we propose a break for the image encryption algorithm proposed in [1].

Chaos based encryption schemes can be broadly divided into two phases: "permutation" and "substitution". Permutation is used to move the image pixels from one position to another, whereas substitution is used make the statistics of the cipher independent on the plaintext. Similarly the algorithm proposed in [1] also constitutes of two parts. First shuffling the original image to disturb the arrangement of pixels by applying Arnold Cat Map [1] several times. Second, a keystream is generated using Chen's chaotic system [3] to mix it with the shuffled pixels to finally obtain the cipher image.

Initially this paper gives a detailed introduction to the cryptosystem, as a basis of the whole letter in Section 2. In Section 3 we describe the method proposed to totally break the system. We also provide an illustrative example followed by the simulation results in Section 4. The paper concludes with final remarks in Section 5.

## 2 Description of the cryptosystem

### 2.1 Arnold Cat Map [1-2]

Assume that we have an N x N image P with pixel co-ordinates I = {(x, y) | x, y = 0, 1, 2… N- 1}.

Arnold cat map is given by

$$\begin{bmatrix} x' \\ y' \end{bmatrix} = A \begin{bmatrix} x \\ y \end{bmatrix} = \begin{bmatrix} 1 & p \\ q & pq+1 \end{bmatrix} \begin{bmatrix} x \\ y \end{bmatrix} \mod N \quad (1)$$

where $p$, $q$ are positive integers and $x'$, $y'$ are the coordinate values of the shuffled pixel. After iterating the map 'n' times we have

$$\begin{bmatrix} x' \\ y' \end{bmatrix} = A^n \begin{bmatrix} x \\ y \end{bmatrix} \mod N = M \begin{bmatrix} x \\ y \end{bmatrix} \mod N \quad (2)$$

where

$$M = \begin{bmatrix} m1 & m2 \\ m3 & m4 \end{bmatrix} = A^n \mod N \quad (3)$$

The shuffled image S is related to the original image P as S(x', y') = P(x, y) where $0 \leq x, y \leq N - 1$.

## 2.2 Chen's chaotic system [3]

Chen's chaotic system is a set of differential equations given as

$$\dot{x} = a(y - x).$$
$$\dot{y} = (c - a)x - xz + cy.$$
$$\dot{z} = xy - bz. \qquad (4)$$

Where a = 35, b = 3 and c € [20, 28.4] for Chen's system to be chaotic.

## 2.3 Encryption Algorithm

Secret keys of the algorithm are parameters p, q, n of Arnold cat map and initial values $x_0$, $y_0$, $z_0$ of Chen's chaotic system. [6]

- Shuffle the image P using Arnold cat map and obtain the shuffled image S. Scan the image S row-by-row and arrange its pixels as sequence S = {$s_1$, $s_2$... $s_{N \times N}$}
- Iterate Chen's chaotic system, obtain the real values $x_i$, $y_i$, $z_i$ where $1 \leq i \leq N_0$ and $N_0$ = (N x N)/3.
- Obtain the key space K = {$k_1$, $k_2$... $k_{N \times N}$} from the key generator.
- Obtain the encrypted system as

  C = {$c_1$, $c_2$... $c_{N \times N}$} as $c_i = s_i \oplus k_i$
  where $\oplus$ represents bitwise OR operation.

## 3 Cryptanalysis of the existing cryptosystem

Here the Intruder intercepts a Cipher Image and has to retrieve the original Plain Image from it. He has on other information available. It is assumed that he has access to the communication channel and the sender or recipient end to gather valuable information which he can utilize later to decrypt the intercepted Cipher Image.

So there are two tasks one to retrieve the Key and the other to obtain the original image from shuffled image

### 3.1 Retrieving the key

According to the Encryption Algorithm the Key is constructed by iterating the Chen's Equation. Since it is not possible to guess the initial parameters, we cannot calculate the particular solution of Chen's Equations and obtain the Key
Hence a method that doesn't depend on reconstructing the Key has to be designed.
As per the definition of Chosen-ciphertext attack , the intruder can choose arbitrary ciphertext and have access to plaintext decrypted from it [16-20].
Let us assume that we have got transient admittance to the encrypting system through Lunch Break Attack.

The task of Encryption machine is to take Input from user and XOR it with the Key and produce the Output. The Key is kept hidden from user.

If a user sends '0' as Input. The following steps take place:

$$\text{Output} = \text{Input} \oplus \text{Key}$$
$$= 0 \oplus \text{Key}$$
$$= \text{Key}$$

Hence unknowingly, the Encryption Machine actually provides the Key as output. This Key can be used to decrypt other messages. We will use this idea in the following section.

### 3.2 Obtaining the Original image from Shuffled Image

Assume that we have an NxN Cipher Image $C = C_1 C_2 C_3 \ldots C_{NxN}$, to decrypt without knowing the secret parameters. Let us again assume that a Lunch Break Attack has been performed. The steps leading to the recovery of the Plain Image P from the intruded Cipher Image C is described below:

**Step 1:** We construct an Image P1 of size 4x4, whose all the elements are 0. For example, let

$$P1 = \begin{pmatrix} 0 & 0 & 0 & 0 \\ 0 & 0 & 0 & 0 \\ 0 & 0 & 0 & 0 \\ 0 & 0 & 0 & 0 \end{pmatrix}$$

Since all elements of Image P1 is 0, all elements of the corresponding Shuffled Image S1 will also be 0.

**Step 2:** We now request the Cipher Image of P1 from the encryption machinery.

Let this Cipher Image be C1

According to the algorithm

$C1 = \text{Key} \oplus \text{Shuffled Image} = \text{Key} \oplus 0 = \text{Key}$

Hence we have obtained the Key without even knowing the secret parameters or method of its construction.

**Step 3:** Now we have already obtained the Key so perform the following steps:

Actual Shuffled Image for Original Plain Image say $S = C \oplus \text{Key}$

Where C is the intercepted Cipher Image

Hence we have obtained our Actual Shuffled Image. Now we simply have to get the Original Image corresponding to this Shuffled Image.

**Step 4:** We next construct another Image P2 of size 4x4, whose elements are 1, 2, 3… 4x4.

$$P2 = \begin{pmatrix} 1 & 2 & 3 & 4 \\ 5 & 6 & 7 & 8 \\ 9 & 10 & 11 & 12 \\ 13 & 14 & 15 & 16 \end{pmatrix}$$

**Step 5:** We now request the Cipher Image C2 of P2 from the encryption machinery.

**Step 6:** Since we have already obtained the Key so perform the following steps:

Shuffled Image for P2 = S2 = C2 $\oplus$ Key

**Step 7:** Now we feed the intercepted Cipher Image C, the obtained Actual Shuffled Image S, Image P2 and its corresponding Shuffled Image S2 to the Analyzer Machine. The Analyzer machine converts the images into their corresponding matrices and does the following jobs:

- Compare the matrices P2 and S2 to understand the shuffling algorithm. Analyzer machine searches for the new position of P2's first element in S2 and similarly repeat the same process for other elements to find their new positions and understand the Shuffling.
- Use the relation to move the elements of S into another matrix P.
- This P is the Original Image we are looking for.

Algorithm 1 gives a detailed description and example of the proposed method. Fig. 1 gives simulation results on a 100x100 image of a sunflower.

## Algorithm 1

Steps to recover the Plain Image P of size 4x4 from the intercepted Cipher Image C using Chosen-ciphertext attack:

PLAIN IMAGE (P)
$$\begin{pmatrix} 23 & 45 & 64 & 32 \\ 179 & 180 & 26 & 58 \\ 67 & 136 & 139 & 20 \\ 17 & 99 & 220 & 100 \end{pmatrix}$$

KEY (K)
$$\begin{pmatrix} 186 & 24 & 39 & 72 \\ 23 & 87 & 47 & 13 \\ 221 & 49 & 50 & 2 \\ 44 & 32 & 65 & 110 \end{pmatrix}$$

CIPHER IMAGE (C)
$$\begin{pmatrix} 174 & 123 & 7 & 252 \\ 6 & 23 & 156 & 134 \\ 240 & 11 & 186 & 102 \\ 54 & 99 & 157 & 121 \end{pmatrix}$$

**Step 1**: We construct an image P1 of size 4x4, whose all elements are 0.

$$P1 = \begin{pmatrix} 0 & 0 & 0 & 0 \\ 0 & 0 & 0 & 0 \\ 0 & 0 & 0 & 0 \\ 0 & 0 & 0 & 0 \end{pmatrix}$$

**Step 2**: We now request the Cipher Image C1 of Image P1 from the encryption machinery.

According to the Encryption Algorithm, the machinery will perform the following operation using the Key i.e. hidden from user.

C1 = Shuffled Image $\oplus$ Key

$$C1 = \begin{pmatrix} 186 & 24 & 39 & 72 \\ 23 & 87 & 47 & 13 \\ 221 & 49 & 50 & 2 \\ 44 & 32 & 65 & 110 \end{pmatrix} \oplus \begin{pmatrix} 0 & 0 & 0 & 0 \\ 0 & 0 & 0 & 0 \\ 0 & 0 & 0 & 0 \\ 0 & 0 & 0 & 0 \end{pmatrix} = \begin{pmatrix} 186 & 24 & 39 & 72 \\ 23 & 87 & 47 & 13 \\ 221 & 49 & 50 & 2 \\ 44 & 32 & 65 & 110 \end{pmatrix}$$

Hence, we obtained the key without even knowing the secret parameter or method of its construction.

**Step 3**: Now since we have already obtained key the key so perform thus we can obtain the original shuffle image.

$$\text{Actual Shuffle Image} = \text{Intercepted Cipher Image} \oplus \text{Key}$$

$$\begin{pmatrix} 174 & 123 & 7 & 252 \\ 6 & 23 & 156 & 134 \\ 240 & 11 & 186 & 102 \\ 54 & 99 & 157 & 121 \end{pmatrix} \oplus \begin{pmatrix} 186 & 24 & 39 & 72 \\ 23 & 87 & 47 & 13 \\ 221 & 49 & 50 & 2 \\ 44 & 32 & 65 & 110 \end{pmatrix} = \begin{pmatrix} 20 & 99 & 32 & 180 \\ 17 & 64 & 179 & 139 \\ 45 & 58 & 136 & 100 \\ 26 & 67 & 220 & 23 \end{pmatrix}$$

**Step 4**: We next construct another image P2 of size 4x4, whose all elements are 1, 2, 3, …. 16.

$$P2 = \begin{pmatrix} 1 & 2 & 3 & 4 \\ 5 & 6 & 7 & 8 \\ 9 & 10 & 11 & 12 \\ 13 & 14 & 15 & 16 \end{pmatrix}$$

**Step 5**: Now let us request the Cipher Image C2 of P2 from encryption machinery

$$C2 = \begin{pmatrix} 182 & 22 & 35 & 78 \\ 26 & 84 & 42 & 6 \\ 223 & 57 & 56 & 18 \\ 43 & 41 & 78 & 111 \end{pmatrix}$$

**Step 6**: Perform the XOR operation of Ciphertext with the key to obtain shuffled image of plain image.

$$\begin{pmatrix} 182 & 22 & 35 & 78 \\ 26 & 84 & 42 & 6 \\ 223 & 57 & 56 & 18 \\ 43 & 41 & 78 & 111 \end{pmatrix} \oplus \begin{pmatrix} 186 & 24 & 39 & 72 \\ 23 & 87 & 47 & 13 \\ 221 & 49 & 50 & 2 \\ 44 & 32 & 65 & 110 \end{pmatrix} = \begin{pmatrix} 12 & 14 & 4 & 6 \\ 13 & 3 & 5 & 11 \\ 2 & 8 & 10 & 16 \\ 7 & 9 & 15 & 1 \end{pmatrix}$$

**Step 7**: Now we need to feed the intercepted image C, the obtained actual shuffled image S, image P2, and its corresponding shuffled image S2 to analyzer machine. The analyzer machine converts the image into their corresponding matrices and does the following job

- Compare the matrices P2 and S2 to understand the shuffling algorithm. Analyzer machine searches for the new position of P2's (1,1) element in S2. From our example P2's (1,1) element has moved to (4,4) in S2. Thus move (4,4) element of S to (1,1) position in another matrix M
- Similarly repeat the process for all other elements to find their new position in the resultant matrix M.
- This M is the original image we are looking for.

$$P2 = \begin{pmatrix} 1 & 2 & 3 & 4 \\ 5 & 6 & 7 & 8 \\ 9 & 10 & 11 & 12 \\ 13 & 14 & 15 & 16 \end{pmatrix} \qquad S2 = \begin{pmatrix} 12 & 14 & 4 & 6 \\ 13 & 3 & 5 & 11 \\ 2 & 8 & 10 & 16 \\ 7 & 9 & 15 & 1 \end{pmatrix}$$

$$M = \begin{pmatrix} 23 & 45 & 64 & 32 \\ 179 & 180 & 26 & 58 \\ 67 & 136 & 139 & 20 \\ 17 & 99 & 220 & 100 \end{pmatrix} \qquad S = \begin{pmatrix} 20 & 99 & 32 & 180 \\ 17 & 64 & 179 & 139 \\ 45 & 58 & 136 & 100 \\ 26 & 67 & 220 & 23 \end{pmatrix}$$

## 4 Results

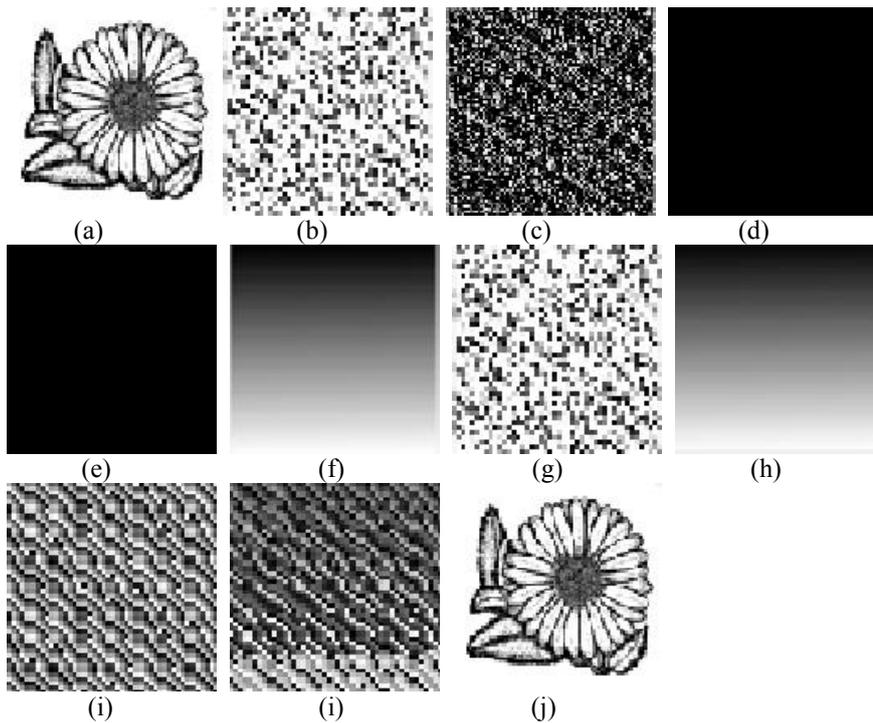

**Fig 1**. (a) Original Plain Image P; (b) Shuffled Image S; (c) Cipher Image C (intercepted by intruder) (d) First Constructed Image P1; (e) Corresponding Shuffled Image S1; (f) Corresponding Cipher Image C1; (g) Actual Shuffled Image calculated;

(h) Second Constructed Image P2; (i) Corresponding Shuffled Image S2; (j) Corresponding Cipher Image C2; (k) Recovered Image M

## 5 Conclusion

In this paper, we have cryptanalysed a recently proposed image encryption scheme by Chosen-ciphertext attack. It is seen that the reuse of the keystream more than once makes it weak against chosen ciphertext attack [5]. The generated keystream is totally independent of the plaintext as well as ciphertext, which makes it unchangeable in every encryption process. The process of shuffling is also predictable and we can extract the original image from the shuffled image. Two couples of plaintext/ciphertext are sufficient to break the system in a Chosen-ciphertext attack scenario. We can make the cryptosystem more secure and harmless to described attack by changing the keystream for every encryption procedure. It can be achieved by either using one time pads as keys or generate the keystream dependent to the plaintext or ciphertext. One time pad [16] is a kind of stream cipher that never reuses its key but in this case distribution of the key is a problem and hence one time pad is impractical in a real secure communication. The solution of making the generation of keystream dependent to the plaintext or the ciphertext is more adequate, secure and practical than one time pads and our future works concerns the same.